\documentclass[letterpaper, 10 pt, conference]{ieeeconf}
\IEEEoverridecommandlockouts
\usepackage{cite}
\usepackage{amsmath,amssymb,amsfonts}
\usepackage{algorithmic}
\usepackage{graphicx}
\usepackage{textcomp}
\usepackage{xcolor}
\def\BibTeX{{\rm B\kern-.05em{\sc i\kern-.025em b}\kern-.08em
    T\kern-.1667em\lower.7ex\hbox{E}\kern-.125emX}}
    
\usepackage{pgfplots}
\usepgfplotslibrary{external}
\usepackage{booktabs}

\newcommand{\ve}[1]{\boldsymbol{#1}}
\newcommand{\vev}[2]{\begin{bmatrix} #1 \\ #2 \end{bmatrix}}

\newcommand{\diff}{\mathrm{d}}
\newcommand{\cair}{c_\mathrm{air}}
\newcommand{\aalpha}{a_{\alpha}}
\newcommand{\aeng}{a_\mathrm{eng}}

\newcommand{\pare}[1]{\left( #1 \right)}

\begin{document}
\bstctlcite{IEEEbib:BSTcontrol}
\title{\LARGE \bf Optimizing Energy-Efficient Braking Trajectories with Anticipatory Road Data for Automated Vehicles}

\author{Andres Alvarez Prado$^{*,1}$, Vladislav Nenchev$^{2}$, Christian Rathgeber$^{2}$%
\thanks{$^{*}$ Corresponding author.}
\thanks{$^{1}$ A. Alvarez Prado is alumnus of the Karlsruhe Institute of Technology. 
{\tt\small andres.alvarez@alumni.kit.edu}}
\thanks{$^{2}$V. Nenchev and C. Rathgeber are with the BMW Group, D-85716 Unterschleissheim, Germany.
{\tt\small \{vladislav.nenchev;christian.rathgeber\}@bmw.de}}%
}

\maketitle

\begin{abstract}
Trajectory planning in automated driving typically focuses on satisfying safety and comfort requirements within the vehicle's onboard sensor range. This paper introduces a method that leverages anticipatory road data, such as speed limits, road slopes, and traffic lights, beyond the local perception range to optimize energy-efficient braking trajectories. For that, coasting, which reduces energy consumption, and active braking are combined to transition from the current vehicle velocity to a lower target velocity at a given distance ahead. Finding the switching instants between the coasting phases and the continuous control for the braking phase is addressed as an optimal trade-off between maximizing coasting periods and minimizing braking effort. The resulting switched optimal control problem is solved by deriving necessary optimality conditions. To facilitate the incorporation of additional feasibility constraints for multi-phase trajectories, a sub-optimal alternative solution based on parametric optimization is proposed. Both methods are compared in simulation. 
\end{abstract}

\begin{keywords}
Trajectory Planning, Switched Systems, Optimal Control, Automated Driving
\end{keywords}

\section{Introduction}
\label{section:introduction}

Adopting an ecologically sustainable driving style, often termed 'eco-driving', exerts a considerable influence on diminishing fuel and energy consumption of passenger vehicles \cite{MIT, bib:miotti}. General guidelines for a human driver to minimize energy consumption can be summarized into shifting up as soon as possible, maintaining a steady speed at the highest gear and low engine revolutions per minute (rpm), anticipating traffic flow, traffic lights, speed limits, and full stops, and maximizing coasting periods \cite{bib:treatise2007,bib:levermore2014,bib:lee2009}.

In the domain of automated driving technology, energy-efficient driving is intrinsically linked with planning and following energy-optimized velocity profiles according to eco-driving principles \cite{bib:freitas2017}. Despite the extensive research and industrial applications of automated eco-driving, several open topics still remain \cite{bib:pan2022}. Notably, the monolithic structure prevalent in many energy-optimal formulations presents a significant computational burden, thereby posing substantial challenges for the
streamlined implementation across a diverse fleet of vehicle models, each with distinct powertrain configurations. 

Energy-efficient longitudinal guidance has often been addressed by formulating an Optimal Control Problem (OCP), where either a fuel or energy consumption model serves as a cost function to be minimized subject to various constraints. The constraints may include control input saturation as well as longitudinal vehicle dynamics. The latter usually considers driving resistances and relies on a detailed powertrain model. Road grade information \cite{bib:ozatay2014} and gear sequences \cite{bib:jing2016} have also been considered within an optimization problem for eco-driving. Online implementations based on analytical solutions obtained with Pontryagin's Maximum Principle (PMP) for typical longitudinal driving maneuvers were proposed in \cite{bib:schwarzkopf1977, bib:saerens2012, bib:sciarretta2015}. However, these approaches require conservative simplifications of the optimization problem to solve the corresponding OCP's analytically. Dynamic programming has also been utilized for energy-optimal motion planning, showcasing its efficacy in handling non-linearities and discrete decisions, such as gear-shifting \cite{bib:hellstrom2010,bib:radke2013,bib:guan2019}. However, these approaches often incur high computational costs.

\begin{figure}
\centering
\includegraphics[width=0.38\textwidth]{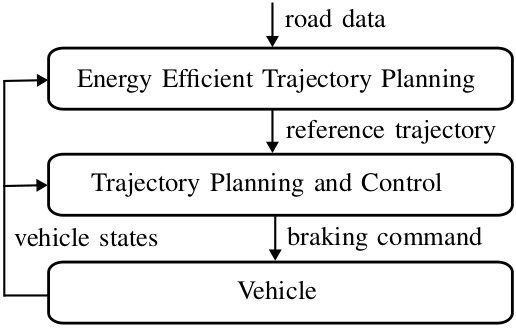}
\caption{Integration of the energy-efficient trajectory planning module into an existing planning and control stack.}
\label{fig:integration}
\end{figure}

Decelerating earlier \cite{bib:miotti} and extending coasting phases \cite{ bib:passenberg2009,bib:yan2023,bib:shakouri} is highly effective at reducing energy consumption. In light of this, our paper introduces a method for the energy-efficient trajectory planning problem during braking maneuvers, specifically focusing on optimizing coasting periods. Our approach addresses eco-driving by maximizing the duration of coasting phases while simultaneously minimizing braking effort, accomplished by solving an OCP. This facilitates a modular architecture: eco-driving is addressed as a separate high-level planning problem, where a detailed powertrain model is not required and its output trajectory can be used as a reference for low-level planning and control approaches, as depicted in Fig.~\ref{fig:integration}. The latter are assumed to be responsible for the safety and comfort requirements of automated driving functions. We focus on the high-level OCP, solved by utilizing the PMP for a longitudinal vehicle model that switches between deceleration trajectories considering coasting and braking phases depending on anticipatory road data. Specifically, the road slope profile is considered to model driving resistances, and upcoming relevant road signs, traffic lights and road curves are interpreted as lower target velocities at a pre-defined distance ahead, i.e., as state constraints. Furthermore, an approximate solution of the OCP is provided based on a parameterized braking phase, which eases the inclusion of additional constraints and yields qualitative similar results to the indirect solution as shown in simulation. Thus, the main contributions of this paper encompass the powertrain-agnostic nature of the OCP formulation, and providing solutions through indirect and direct approaches possible through readily available computationally efficient solvers.

The paper is organized as follows: In Sec.~\ref{section:problem_formulation} the vehicle dynamics model and the OCP formulation are introduced. Then, the resulting switched OCP is solved by means of the indirect method in Sec.~\ref{section:indirect_approach}. In Sec.~\ref{section:direct_approach_approximate_solution}, an approximate parametric solution for the switched OCP is presented. Then, simulation results for a braking scenario are shown and both methods are compared in Sec.~\ref{section:case_study}.  

\section{Problem Formulation}
\label{section:problem_formulation}

Assuming low lateral accelerations, only the longitudinal dynamics of the vehicle have to be considered. The continuous state of the system $\ve{x}$ evolving over time $t$ contains the traveled distance $s$ and the vehicle velocity $v$, i.e., $\ve{x}(t) = \left[ s(t), v(t) \right]^\intercal$ and $\ve{x}(t) \subset X \in \mathbb{R}^2_{\geq 0}$. Let $F_b(t)$ denote the braking force, and $F_{\mathrm{res}}(s,v)$ -- the force acting on the vehicle due to the driving resistances depending on the current $s$ and $v$. Then, the longitudinal dynamics are described as
\begin{align*}
\begin{aligned}
\dot{s}(t) &= v(t), \\
\dot{v}(t) &= -\frac{1}{m} \bigl(F_b(t) + F_{\mathrm{res}}\pare{s(t),v(t)} \bigr),
\end{aligned}
\end{align*}
where $m$ is the overall vehicle mass. The driving resistance force $F_{\mathrm{res}}(s,v)$ takes into account rolling friction, road slope, air, and engine drag resistance forces, and reads
\begin{multline*}
F_{\mathrm{res}}(s,v) = c_r m g \cos \left( \alpha(s) \right) + m g \sin \left( \alpha(s) \right) \\
+ \frac{1}{2} \rho c_d A_f v^2(t) + F_{\mathrm{drag}},
\end{multline*}
where $c_r$ is the rolling friction coefficient, $g$ is the gravitational acceleration, $\alpha(s)$ is the road slope angle varying over the distance, $\rho$ is the air density, $c_d$ is the aerodynamic coefficient of drag, $A_f$ is the cross-sectional area of the vehicle's frontal surface, and $F_{\mathrm{drag}}$ is the engine drag resistance force. The latter models the coasting phases by 
\begin{equation*}
F_{\mathrm{drag}} =
	\begin{cases}
		F_{\mathrm{eng}}, & \text{engaged coasting} \\
		0, & \text{disengaged coasting}
	\end{cases}
\end{equation*}
where $F_{\mathrm{eng}}$ refers to a feedback or estimated value from the vehicle's engine drag potential force. Note that the vehicle model does not account for predictive gear-shifting behaviour, but reduces gear-shifting into engaged or disengaged coasting phases, where the braking force $F_b(t)$ is set to zero. In case of hybrid or electric vehicles, $F_{\mathrm{eng}}$ models the current recuperation force of the electric powertrain.

\begin{figure}
\centering
\includegraphics[width=0.47\textwidth]{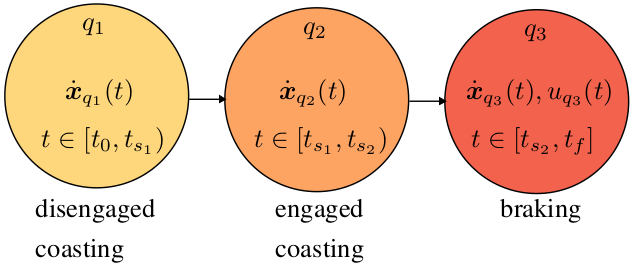}
\caption{A switched system for anticipating braking maneuvers.}
\label{fig:switched_system}
\end{figure}

Due to the discrete nature of the coasting and braking phases and the difference in the continuous state evolution within these phases, the longitudinal behavior of the vehicle is modeled as a switched system. Let $q_1$ denote the disengaged coasting, $q_2$ -- the engaged coasting, and $q_3$ -- the braking mode. Then, the set of possible discrete modes is given by $Q = \{ q_1, q_2, q_3 \}$. Let the control input depending on the discrete mode $q$ be given by
\begin{equation}
\label{eq:discrete_control}
u(t,q) =
	\begin{cases}
		0, & q=q_1 \\
		- a_{\mathrm{eng}}, & q=q_2 \\
		u_{q_3}(t), & q=q_3
	\end{cases}
\end{equation}
where $a_{\mathrm{eng}} = F_{\mathrm{eng}}/m$ is the vehicle's engine drag or recuperation deceleration, and $u_{q_3}(t) = - F_b(t)/m$ is the deceleration command to the system in case of an active braking phase, where $u_{q_3}(t)\in U\subset \mathbb{R}_{\leq 0}$. Assuming a constant road slope angle $\alpha(s) \equiv \alpha$ and using \eqref{eq:discrete_control}, the simplified longitudinal dynamics $\ve{\dot{x}}_q(t) =\ve{f}_{q} \left( \ve{x}_q(t), u(t,q) \right)$ are described by
\begin{equation}
\label{eq:hybrid_model}
\ve{\dot{x}}_q(t) = \vev{v(t)}{- c_{\mathrm{air}} v^2(t) - \aalpha + u(t,q)},
\end{equation}
for $q \in Q$, with the air resistance coefficient $c_{\mathrm{air}} = \frac{1}{2m} \rho c_d A_f$ and $\aalpha = c_r g \cos \left( \alpha \right) + g \sin \left( \alpha \right)$ modeling the rolling friction and gradient resistance deceleration. Since $a_{\mathrm{eng}}$ is considered to be a feedback value from the vehicle's actuators potential, it is held constant throughout the state evolution in time. Therefore, both coasting phases, $q_1$ and $q_2$, can be considered as autonomous systems, whereas the active braking phase $q_3$ is the only controlled sub-system within our problem formulation. 
From a comfort perspective, arranging the phases in ascending order with respect to the absolute deceleration experienced by the passengers in the vehicle minimizes the magnitude of the jerk impulses while switching phases. Thus, the vehicle should start in the disengaged coasting phase $q_1$ at the initial state ${\ve{x}_0 = \left[s_0, v_0 \right]^\intercal \in \mathbb{R}^2_{\geq0}}$, and then optionally switch to engaged coasting $q_2$. To mitigate plant-model mismatches or disturbances, it is essential to approach the target states  ${\ve{x}_{f} = \left[ s_f, v_f \right]^\intercal \in \mathbb{R}^2_{\geq 0}}$ within the controlled phase $q_3$. The switched system is depicted schematically in Fig.~\ref{fig:switched_system}. The trajectory starts at the initial time $t_0$ and ends at the final time $t_f$. The switching times $t_{s_1}, t_{s_2}$ between the modes are free to choose and no state jumps should occur upon switching instants, i.e.  $\ve{x}_{q_1}(t_{s_1}) = \ve{x}_{q_2}(t_{s_1})$ and $\ve{x}_{q_2}(t_{s_2}) = \ve{x}_{q_3}(t_{s_2})$ holds.
Penalizing energy consumption can be achieved by the well-known eco-driving heuristics of minimizing braking effort and maximizing coasting periods. Thus, the cost function is composed of a cost term $J_u$ that penalizes the control input energy while braking and a cost term $J_t$ that penalizes the overall duration of the trajectory. The total cost is given by 
\begin{equation}
\label{eq:cost}
J = J_u + J_t = \frac{w_u}{2} \int_{t_{s_2}}^{t_f} u_{q_3}^2(t) \diff t + w_t \int_{t_0}^{t_f} \diff t,
\end{equation}
where $w_u \in \mathbb{R}_{>0}$ and $w_t \in \mathbb{R}_{>0}$ are the corresponding weights for each term. While $J_u$ denotes the cost of braking effort, $J_t$ favours longer coasting periods. During the coasting phases no active braking is applied and the vehicle velocity decrease results from the driving resistances. Therefore, the vehicle velocity is kept relatively high in comparison to an active braking phase and the overall duration is reduced when using longer coasting periods.

By choosing the switching time instants $t_{s_1}$ and $t_{s_2}$, as well as the braking deceleration within the braking phase, denoted by $u_{q_3}(t)$, the goal is to minimize the combined costs \eqref{eq:cost} with respect to \eqref{eq:hybrid_model} while moving from an initial state $\ve{x}_0$ to the target state $\ve{x}_f$. Thus, the OCP
\begin{subequations}
\label{eq:switched_ocp}
\begin{align}
&\underset{t_{s_1}, t_{s_2}, u_{q_3}(t)}{\min} \quad J_u + J_t \label{eq:ocp:cost}\\
\text{s.t.}&  \quad \ve{\dot{x}}_q(t) =\ve{f}_{q} \left( \ve{x}_q(t), u(t,q) \right),\ q \in Q,  \label{eq:ocp:switched_system} \\
& \quad \ve{x}_{q_1}(t_0) = \ve{x}_0,  \label{eq:ocp:initial_condition}\\
& \quad \ve{x}_{q_2}(t_{s_1}) = \ve{x}_{q_1}  (t_{s_1}),\label{eq:ocp:first_switch} \\
& \quad \ve{x}_{q_3}(t_{s_2}) = \ve{x}_{q_2}(t_{s_2}), \label{eq:ocp:second_switch}\\
& \quad \ve{x}_{q_3}(t_f) = \ve{x}_f,\label{eq:ocp:final_condition}\\
& \quad u_\mathrm{min} \leq u_{q_3}(t) \leq 0, \label{eq:ocp:control_input}\\
& \quad 0 \leq t_0 \leq t_{s_1} \leq t_{s_2} \leq t_f, \label{eq:ocp:time_conditions}
\end{align}
\end{subequations}
where $u_{\mathrm{min}}\in\mathbb{R}_{<0}$ denotes the minimal allowed deceleration and the final time $t_f$ is free, is addressed in the following.

\section{Indirect Approach}
\label{section:indirect_approach}

The first approach to solve the switched OCP \eqref{eq:switched_ocp} is based on applying the Hybrid Minimum Principle (HMP) presented in \cite{bib:caines} to derive necessary optimality conditions. For switched systems without state jumps, applying the HMP results in similar optimality conditions as the PMP for each individual phase, with additional continuity conditions at the switching instants.

\subsection{Formulation of the Hamiltonians}
\label{subsection:formulation_of_hamiltonians}

With the costates defined as $\ve{\lambda}_{q}(t) = \left[ \lambda_{s_{q}}(t) , \lambda_{v_{q}}(t) \right]^{\intercal}$, the family of system Hamiltonians reads
\begin{multline}
\label{eq:family_hamitlonians}
H_{q}(\ve{x}_{q}(t), \ve{\lambda}_{q}(t), u_q(t)) = \ve{\lambda}_{q}^{\intercal}(t) \ve{f}_{q}\pare{\ve{x}_{q}(t), u_q(t)} \\+ l_q(u_q(t)),
\end{multline}
for $q \in Q$, and where $l_q(u_q(t))$ denotes the family of cost functions given by
\begin{align*}
l_q(u_q(t)) =
	\begin{cases}
		w_t, & q=q_1, q_2 \\
		\frac{w_u}{2} u_{q_3}^2(t) + w_t, & q=q_3.
	\end{cases}
\end{align*}
Based on the HMP, the Hamiltonians must be continuous at the switching instants
\begin{align}
H_{q_2}(t_{s_1}) &= H_{q_1}(t_{s_1}), \label{eq:H_ts1} \\
H_{q_3}(t_{s_2}) &= H_{q_2}(t_{s_2}), \label{eq:H_ts2}
\end{align}
and the free final time condition on the braking phase
\begin{equation}
\label{eq:H_tf}
H_{q_3}(t_f) = 0
\end{equation}
has to be satisfied. Further, the optimal control input that minimizes \eqref{eq:family_hamitlonians} at $q_3$ needs to fulfill the stationary condition $\partial H_{q_3} / \partial u_{q_3} \big|_{(\cdot)^*} = 0$, yielding the optimal continuous control
\begin{equation}
\label{eq:opt_control}
u_{q_3}^*(t) = - \frac{1}{w_u} \lambda_{v_{q_3}}^*(t).
\end{equation}
\subsection{Costate dynamics}
\label{subsection:costate_dynamics}
The costate dynamics at each phase are described by
\begin{equation*}
\dot{\ve{\lambda}}_{q}^*(t) = -\frac{\partial H_{q}}{\partial \ve{x}_{q}}\bigg|_{(\cdot)^*} = \vev{0}{-\lambda_{s_{q}}^*(t) + 2\cair v_{q}^*(t)\lambda_{v_{q}}^*(t)},
\end{equation*}
for $q \in Q$, with the transition conditions
\begin{align}
\ve{\lambda}_{q_2}^*(t_{s_1}) &= \ve{\lambda}_{q_1}^*(t_{s_1}), \label{eq:costate_ts1}\\
\ve{\lambda}_{q_3}^*(t_{s_2}) &= \ve{\lambda}_{q_2}^*(t_{s_2}). \label{eq:costate_ts2}
\end{align}
Note that since $\dot{\lambda}_{s_{q}}^*(t) = 0$ for all $q \in Q$ and conditions \eqref{eq:costate_ts1}, \eqref{eq:costate_ts2} hold, the costate related to traveled distance is constant throughout the entire trajectory, i.e., $\lambda_{s_{q}}^*(t) \equiv \lambda_s$ for all $q \in Q$. Thus, the costate dynamics are reduced to
\begin{equation}
\label{eq:costate_dynamics}
\dot{\lambda}_{v_{q}}^*(t) = - \lambda_s + 2 \cair v^*_{q_1}(t) \lambda_{v_{q_1}}^*(t).
\end{equation}
Inserting \eqref{eq:opt_control} and \eqref{eq:ocp:final_condition} into \eqref{eq:H_tf} and solving for ${\lambda_{v_{q_3}}^*(t_f) = \lambda_{v}(t_f)}$, the following condition on the final costate value arises
\begin{multline}
\label{eq:costate_v_tf}
\lambda_{v}(t_f) = - w_u \pare{\cair v_f^2 + \aalpha} \\
+ \sqrt{ w_u^2 \pare{\cair v_f^2 + \aalpha}^2 + 2 w_u (w_t + \lambda_s v_f)}.
\end{multline}
In addition, plugging \eqref{eq:ocp:second_switch} and \eqref{eq:costate_ts2} into \eqref{eq:H_ts2}, the following condition on the costate at the second switching instant ${\lambda^*_{v_{q_3}}(t_{s_2}) = \lambda^*_{v_{q_2}}(t_{s_2}) = \lambda_{v}(t_{s_2})}$ should be fulfilled
\begin{equation}
\label{eq:costate_v_ts2}
\lambda_{v}(t_{s_2}) = 2 w_u \aeng.
\end{equation}
Analogously, inserting \eqref{eq:ocp:first_switch} and \eqref{eq:costate_ts1} into \eqref{eq:H_ts1} yields the transition condition for $\lambda^*_{v_{q_2}}(t_{s_1}) = \lambda^*_{v_{q_1}}(t_{s_1}) = \lambda_{v}(t_{s_1})$:
\begin{equation}
\label{eq_lambda_ts1}
\lambda_{v}(t_{s_1}) = 0.
\end{equation}
\subsection{State dynamics}
\label{subsection:continuous_state_dynamics}
Since both coasting phases are autonomous systems, their state evolution is independent of their respective costates. However, for the braking phase, the optimal state evolution is given by using \eqref{eq:opt_control} as the optimal control input  
\begin{equation}
\label{eq:opt_braking_dynamics}
\ve{\dot{x}}_{q_3}^*(t) = \vev{v_{q_3}(t)}{- \cair v^2(t) - \aalpha - \frac{1}{w_u} \lambda_{v_{q_3}}^*(t)}.
\end{equation}
\subsection{Solving the switched optimal control problem}
\label{subsection:solving_the_switched_ocp}
To solve the switched OCP, we make use of the fact that the state evolution from the autonomous coasting phases are decoupled from their corresponding costates. Since a closed-form solution for the state dynamics at the coasting phases exists, the costates are described by non-homogeneous, first-order Ordinary Differential Equations (ODEs) with time-varying coefficients. These are also amenable for a closed-form solution. The corresponding closed-form expressions for the state evolutions and costates are given in Appendices \ref{appendix:closed_form_state_evolution_coasting_phases} and \ref{appendix:closed_form_adjoint_process_coasting_phases}.

The optimal braking phase trajectory is described by the non-linear coupled ODEs given in \eqref{eq:opt_braking_dynamics} and \eqref{eq:costate_dynamics}, whose solution relies on numeric integration. By utilizing the closed-form expressions of both coasting phases, we can reduce the switched OCP into a single Boundary-Value Problem (BVP) described by the state dynamics at braking and its corresponding costate related to the velocity state. Thus, we define ${\ve{z}(t) = \left[ s_{q_3}^*(t) , v_{q_3}^*(t) , \lambda_{v_{q_3}}^*(t) \right]^{\intercal} \in Z \subset \mathbb{R}^{3}_{\geq 0}}$ as
\begin{equation}
\label{eq:original_bvp}
\dot{\ve{z}}(t) 
= \begin{bmatrix} v_{q_3}^*(t) \\
- \cair v_{q_3}^{*2}(t) - \aalpha - \frac{1}{w_u} \lambda_{v_{q_3}}^*(t) \\
- \lambda_s + 2 \cair v^*_{q_3}(t) \lambda_{v_{q_3}}^*(t)
\end{bmatrix},\ t \in \left[t_{s_2}, t_f \right]
\end{equation}
with ${\ve{z}(t_{s_2})
= \left[ s_{q_2}^*(t_{s_2}), v_{q_2}^*(t_{s_2}), \lambda_{v}(t_{s_2}) \right]^\intercal}$ and ${\ve{z}(t_{f}) = \left[ s_f,v_f,\lambda_{v}(t_{f}) \right]^\intercal}$ as boundary conditions.
In order to solve \eqref{eq:original_bvp} with an off-the-shelf BVP solver, we transform the original free end time problem into a fixed end time configuration via time variable transformation given by ${t = (t_f - t_{s_2}) \tau + t_{s_2}}$, with ${\tau \in \left[ 0, 1\right]}$ and
\begin{equation*}
\frac{\diff \pare{\cdot}}{\diff t}  = \frac{\diff \pare{\cdot} }{\diff \tau} \frac{\diff \tau}{\diff t} = \frac{\diff \pare{\cdot} }{\diff \tau} \frac{1}{t_f - t_{s_2}}.
\end{equation*}
Thus, we define $\ve{\xi}(\tau) := \ve{z}\pare{(t_f - t_{s_2}) \tau + t_{s_2}} \in \Xi \subset \mathbb{R}^{3}_{\geq 0}$, with dynamics
\begin{equation*}
\frac{\diff \ve{\xi}(\tau) }{\diff \tau} = \ve{h}( \ve{\xi}(\tau)) = (t_f - t_{s_2}) \dot{\ve{z}}(t),\ \tau \in \left[ 0, 1\right],
\end{equation*}
the initial condition $\ve{\xi}(0) = \ve{z}(t_{s_2})$, and the final condition $\ve{\xi}(1) = \ve{z}(t_{f})$.
The switching and final times are gathered in a vector of unknown parameters ${\ve{t} = \left[ t_{s_1} , t_{s_2} , t_f \right]^{\intercal} \in T \subset \mathbb{R}^{3}_{\geq 0}}$, which are to be found along $\ve{\xi}(\tau)$. We describe the still unknown costate $\lambda_s$ in dependency of $\ve{t}$ by evaluating \eqref{eq:costate_q2_closed_form} at $t_{s_1}$. Finally, the BVP reads
\begin{subequations}
\label{eq:final_bvp}
\begin{align}
& \text{Find} \quad \ve{\xi}(\tau), \ve{t} \\
\text{s.t.}& \quad  \frac{\diff {\ve{\xi}}(\tau)}{\diff \tau} = \ve{h}( \ve{\xi}(\tau),\ve{t}), \quad \tau \in \left[ 0, 1\right] \\
& \quad  \ve{\xi}(0) = \ve{b}_0(\ve{t}), \\
& \quad  \ve{\xi}(1) = \ve{b}_f(\ve{t}),
\end{align}
\end{subequations}
with $ \ve{b}_0(\ve{t}) = \ve{z}(t_{s_2})$ and $ \ve{b}_f(\ve{t}) = \ve{z}(t_{f})$. 
\section{Direct Approach for approximate solution}
\label{section:direct_approach_approximate_solution}
In the following, a parametric optimization approach is presented as an alternative to the indirect approach described in the previous section. An advantage of using a parametric approach is the easier extendability with additional constraints, which were previously neglected within the indirect method. Control input saturation constraints \eqref{eq:ocp:control_input} are important to avoid high decelerations which result in uncomfortable trajectories and might not be feasible to realize by the vehicle actuators. In addition, constraints on the discrete switching time instants \eqref{eq:ocp:time_conditions} avoid negative duration of the discrete phases, guaranteeing feasible switching and final times.
Motivated by the optimal trajectories obtained from the indirect approach --- shown later in Sec.~\ref{section:case_study} --- we approximate the optimal control input $u_{q_3}^*(t)$ with a state feedback control law described by 
\begin{equation}
\label{eq:control_state_feedback}
\hat{u}_{q_3}(t) = - u_m v_{q_3}(t) + u_n,
\end{equation}
with control parameters $u_m \in \mathbb{R}$ and $u_n \in \mathbb{R}$. The closed-form solution in time and space domain of the braking dynamics with a control input from type \eqref{eq:control_state_feedback}, as well as the analytical expression for the braking phase duration, can be found in Appendix~\ref{appendix:closed_form_state_evolution_braking_phase}. By inserting \eqref{eq:velocity_state_feedback_closed_form} and \eqref{eq:braking_duration_state_feedback_closed_form} into the cost function \eqref{eq:cost} the costs can be reformulated in the dependency of the switching times and control input parameters.

Note that the state evolution expressions of the coasting phases $q_1$ and $q_2$ --- including initial and transition conditions on the states --- are encoded into the closed-form expression of the state evolution in the braking phase $q_3$, which has been inserted into the cost function. Hence, \eqref{eq:ocp:first_switch} --- \eqref{eq:ocp:second_switch} are always fulfilled and can be neglected. The final condition on the states \eqref{eq:ocp:final_condition} is reformulated in the space domain by defining the equality constraint ${g_f = s_f - s_{q_3}(v_f) = 0}$, where $s_{q_3}(v_{q_3})$ is given in \eqref{eq:distance_velocity_state_feedback_closed_form}. The saturation on the continuous control input is considered at critical times by ${\ve{g}_{u_\mathrm{max}} = \left[-\hat{u}_{q_3}(t_{s_2}),-\hat{u}_{q_3}(t_{f})\right]^\intercal \geq \boldsymbol{0}}$ and ${\ve{g}_{u_\mathrm{min}} = \left[ - u_\mathrm{min} + \hat{u}_{q_3}(t_{s_2}),- u_\mathrm{min} + \hat{u}_{q_3}(t_{f}) \right]^\intercal  \geq \boldsymbol{0}}$. An additional constraint ${g_c = u_m^2 - 4 \cair \pare{\aalpha - u_n} \geq 0}$ is added to account for feasibility of the closed-form expressions of the braking phase. The condition on strictly increasing critical times \eqref{eq:ocp:time_conditions} can be simplified using the time transformation $\Delta t_{q_1} = t_{s_1} - t_0$, $\Delta t_{q_2} = t_{s_2} - t_{s_1}$, and $\Delta t_{q_3} = t_f - t_{s_2}$ throughout the entire optimization problem. Finally, by defining the parameter vector ${\ve{\theta} = \left[ \Delta t_{q_1} , \Delta t_{q_2} , u_m , u_n \right]^{\intercal} \in \mathbb{R}^4}$, we approximate the original switched OCP by the Nonlinear Program (NLP)
\begin{subequations}
\label{eq:nlp}
\begin{align}
\underset{\ve{\theta}}{\min}&  \quad J \pare{\ve{\theta}}   \\
\text{s.t.} & \quad g_{\mathrm{eq}}\pare{\ve{\theta}} = 0, \\
& \quad \ve{g}_{\mathrm{ineq}}\pare{\ve{\theta}} \geq \boldsymbol{0}, \label{eq:nlp_ineq}
\end{align}
\end{subequations}
with the equality constraint $g_{\mathrm{eq}}\pare{\ve{\theta}} = g_f$ and the inequality constraint vector ${\ve{g}_{\mathrm{ineq}}\pare{\ve{\theta}} = \left[\ve{g}_{u_\mathrm{max}}, \ve{g}_{u_\mathrm{min}}, g_c, \theta_1, \theta_2 \right]^\intercal}$.

\section{Case Study}
\label{section:case_study}
The switched OCP is solved in a simulation example using the indirect and direct method. The indirect approach yields the BVP \eqref{eq:final_bvp}, which is solved using the BVP subroutine of the SciPy Python library \cite{bib:bvp_ascher}. The direct approach yields the NLP \eqref{eq:nlp} solved using the Interior Point Optimizer (IPOPT) \cite{bib:ipopt}. A representative parameter set given in Tab.~\ref{table:case_study_parameters} is used. The switched OCP plans a multi-phase trajectory which decelerates the vehicle from an initial velocity of 150 km/h to 100 km/h in 500 m. 

\begin{table}
\caption{Parameters used in case study.}
	\centering
	\begin{tabular}{llll}
		\toprule
		\textbf{Parameter} & \textbf{Symbol} & \textbf{Value} & \textbf{Unit} \\
		\cmidrule(r){1-1}\cmidrule(lr){2-2}\cmidrule(lr){3-3}\cmidrule(l){4-4} 
		Vehicle mass & $m$ & 2795 & kg \\
		Cross-sectional area & $A_f$ & 2.26 & $\text{m}^2$ \\
		Engine drag deceleration & $a_{\mathrm{eng}}$ & 0.4 & m/$\text{s}^2$ \\
		Coefficient of drag & $c_d$ & 0.25 & -- \\
		Rolling coefficient of friction & $c_r$ & 0.015 & -- \\
		\midrule
		Road slope angle & $\alpha$ & 2 & deg \\
		Gravitational acceleration & $g$ & 9.81 & m/$\text{s}^2$ \\
		Air density & $\rho$ & 1.29 & kg/$\text{m}^3$ \\
		\midrule
		Weight on overall trajectory duration & $w_t$ & 1.0 & -- \\
		Weight on control input energy & $w_u$ & 0.1 & -- \\
		\midrule
		Control input constraint & $u_{\mathrm{min}}$ & -2.0 & m/$\text{s}^2$ \\
		\bottomrule
	\end{tabular}
\label{table:case_study_parameters}
\end{table}

\begin{figure}
	\centering
	\includegraphics[width=0.38\textwidth]{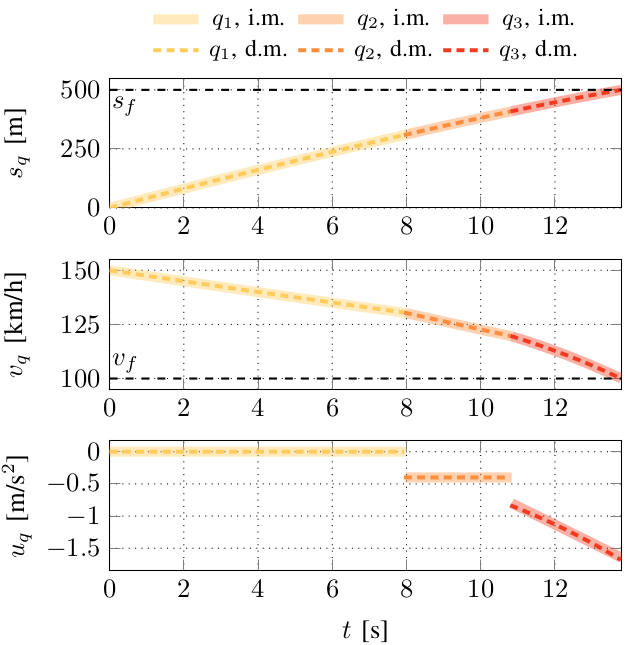}
	\caption{Comparison of optimal multi-phase trajectories computed with the indirect method (i.m.) and its approximation using the direct method (d.m.).}
	\label{fig:results}
\end{figure}

Fig.~\ref{fig:results} shows the result of the simulation example. Both methods show similar qualitative results. The indirect approach results in the optimal phase durations ${\pare{\Delta t_{q_1}^{\mathrm{i.m.}}, \Delta t_{q_2}^{\mathrm{i.m.}}, \Delta t_{q_3}^{\mathrm{i.m.}}} \approx \pare{7.98\ \mathrm{s}, 2.86\  \mathrm{s}, 2.95\  \mathrm{s}}}$, whereas the phase durations with the direct approach are given by ${\pare{\Delta t_{q_1}^{\mathrm{d.m.}}, \Delta t_{q_2}^{\mathrm{d.m.}}, \Delta t_{q_3}^{\mathrm{d.m.}}} \approx \pare{7.93\ \mathrm{s}, 2.87\ \mathrm{s}, 2.98\ \mathrm{s}}}$. In addition, the optimal state feedback control parameters are ${u_m \approx - 1.55 \cdot 10^{-1}\ \mathrm{s}^{-1}}$ and ${u_n \approx -5.99\ \mathrm{m}/\mathrm{s}^2}$. The cost from the indirect approach trajectory, $J^{\mathrm{i.m.}} \approx 14.01588$, is slightly lower than its direct counterpart, $J^{\mathrm{d.m.}} \approx 14.01591$. The similarity of both trajectories highlights the good approximation of the optimal braking control given by the control law from \eqref{eq:control_state_feedback}. Note that more than half of the traveled distance is covered by disengaged coasting, followed by a shorter engaged coasting phase. Finally, the target states are reached with a  moderate braking deceleration command within the braking phase. Thus, the relevant eco-driving principles for braking maneuvers are well represented by the stated OCP, potentially enhancing the energy-saving capabilities of automated driving vehicles. Note that solving the OCP in a single shot optimization was sufficient to reach the target state conditions in simulation. In a practical realization, the OCP will be solved repetitively in a receding horizon fashion to consider updated road information and to account for disturbances. 

\section{Conclusion and Outlook}
\label{section:conclusion_and_outlook}

We proposed two solutions for the Optimal Control Problem (OCP) associated with planning energy-efficient longitudinal braking trajectories with initial and end constraints based on eco-driving principles and without requiring complex fuel or energy consumption models. Using an indirect approach, the original switched OCP was transformed into a single boundary value problem. To ease the extension of the OCP with additional constraints, a direct approach that translates the original OCP into a parametric optimization problem was also presented. The algorithm's powertrain-agnostic nature and its compatibility with off-the-shelf numerical solvers make it particularly suitable for integration into diverse vehicular trajectory planning and control architectures. In the context of electric vehicles, the method can be readily adapted by excluding the disengaged coasting phase and considering the engaged coasting mode as regenerative braking. 

Future work will explore the incorporation of variable road slope data. In addition, more realistic simulation scenarios and real-world vehicle testing should be considered to assess the energy-saving achieved with the proposed methods.

\appendices

\section{Closed-form state evolution for coasting phases}
\label{appendix:closed_form_state_evolution_coasting_phases}
For the coasting phases, the closed-form solution of \eqref{eq:hybrid_model} is given by
\begin{equation}
\label{eq:coasting_velocity_closed_form}
v_{q}(t) = b_{q,1} \tan \left( b_{q,2} \left(t - t_{q,0} \right) + \arctan \left( \frac{v_{q,0}}{b_{q,1}} \right)    \right), 
\end{equation} 
and
\begin{equation*}
s_{q}(t)  = s_{q,0} + \frac{1}{2 c_\mathrm{air}} \ln \left( \frac{ \left( \frac{v_{q,0}}{b_{q,1}} \right)^2 + 1}{ \left( \frac{v_{q}(t)}{b_{q,1}} \right)^2 + 1} \right)
\end{equation*}
for $q \in \{q_1, q_2\}$. Note that the initial and transition conditions are encoded into the closed-form expressions as $(s_{q_1,0}, v_{q_1,0}) = (s_{0}, v_{0})$ at $t_{q_1,0} = t_0$ for the disengaged coasting phase, and $(s_{q_2,0},v_{q_2,0}) = (s_{q_1}(t_{s_1}),v_{q_1}(t_{s_1}))$ at $t_{q_2,0} = t_{s_1}$ for the engaged coasting phase. Furthermore, we define ${b_{q_1,1} = \sqrt{\aalpha / \cair}}$, ${b_{q_1,2} = - \sqrt{\aalpha \cair}}$, ${b_{q_2,1} = \sqrt{\pare{\aalpha + a_{\mathrm{eng}}} / \cair}}$, and ${b_{q_2,2} = - \sqrt{ \pare{\aalpha + a_{\mathrm{eng}}} \cair}}$ as constant terms.

\section{Closed-form costates for coasting phases}
\label{appendix:closed_form_adjoint_process_coasting_phases}
Inserting \eqref{eq:coasting_velocity_closed_form} in \eqref{eq:costate_dynamics} and considering the transition conditions from \eqref{eq:costate_ts1} and \eqref{eq:costate_ts2}, the closed-form expressions for the coasting phases are given by 
\begin{multline}
\label{eq:costate_q2_closed_form}
\lambda_{v_{q_2}}^*(t) = \lambda_s \frac{v_{q_2}^*(t) - v_{q_2}^*(t_{s_2})}{\cair {v_{q_2}^*}^2(t) + \aalpha + a_{\mathrm{eng}}} \\
 + \lambda_{v}(t_{s_2}) \frac{\cair {v_{q_2}^*}^2(t_{s_2}) + \aalpha + a_{\mathrm{eng}} }{\cair {v_{q_2}^*}^2(t) + \aalpha + a_{\mathrm{eng}}}
\end{multline}
for the engaged coasting phase, and
\begin{equation*}
\label{eq:costate_q1_closed_form}
\lambda_{v_{q_1}}^*(t) = \lambda_s \frac{v_{q_1}^*(t) - v_{q_1}^*(t_{s_1})}{\cair {v_{q_1}^*}^2(t) + \aalpha} + \lambda_{v}(t_{s_1}) \frac{\cair {v_{q_1}^*}^2(t_{s_1}) + \aalpha}{\cair {v_{q_1}^*}^2(t) + \aalpha},
\end{equation*}
for the disengaged coasting phase.
\section{Closed-form braking phase}
\label{appendix:closed_form_state_evolution_braking_phase}
In the time domain, the closed-form solution of the braking phase \eqref{eq:hybrid_model} using the state feedback control \eqref{eq:control_state_feedback} is given by
\begin{multline}
\label{eq:distance_state_feedback_closed_form}
s_{q_3}(t) = s_{q_2}(t_{s_2}) + \frac{b_{q_3,1} - u_m}{2 \cair} \pare{t - t_{s_2}} \\
 + \frac{1}{\cair} \ln \pare{ \frac{1 - b_{q_3,2} \exp \pare{- b_{q_3,1} \pare{t - t_{s_2} } }}{1 - b_{q_3,2}} }
\end{multline}
and 
\begin{multline} 
\label{eq:velocity_state_feedback_closed_form}
v_{q_3}(t) = \frac{\pare{u_m + b_{q_3,1}} b_{q_3,2} \exp \pare{- b_{q_3,1} \pare{t - t_{s_2} } }}{2 \cair \pare{1 - b_{q_3,2} \exp \pare{- b_{q_3,1} \pare{t - t_{s_2} } }} } \\
- \frac{\pare{u_m - b_{q_3,1}}}{2 \cair \pare{1 - b_{q_3,2} \exp \pare{- b_{q_3,1} \pare{t - t_{s_2} } }} },
\end{multline}
with 
\begin{align*}
b_{q_3,1}&=\sqrt{u_m^2 - 4 \cair \pare{\aalpha - u_n} },\\
b_{q_3,2}&= \frac{2 \cair v_{q_2}(t_{s_2}) + u_m - b_{q_3,1}}{2 \cair v_{q_2}(t_{s_2}) + u_m + b_{q_3,1}}.
\end{align*}
Integrating in the space domain via $\diff v / \diff t = (\diff v / \diff s) v$, we obtain the expression for the braking phase:
\begin{multline}
\label{eq:distance_velocity_state_feedback_closed_form}
s_{q_3}(v_{q_3}) = s_{q_2}(t_{s_2}) \\
+ \frac{1}{2 \cair} \ln \pare{ \frac{\cair v_{q_2}^2(t_{s_2}) + u_m v_{q_2}(t_{s_2})  + \aalpha - u_n}{\cair v_{q_3}^2 + u_m v_{q_3} + \aalpha - u_n}} \\
+ \frac{u_m}{2 \cair b_{q_3,1}} \left( \ln \left( \frac{2 \cair v_{q_3} + u_m - b_{q_3,1}}{2 \cair v_{q_3} + u_m + b_{q_3,1}} \right) \right. \\
+ \left. \ln \left( \frac{2 \cair v_{q_2}(t_{s_2}) + u_m + b_{q_3,1}}{2 \cair v_{q_2}(t_{s_2}) + u_m - b_{q_3,1}} \right) \right).
\end{multline}
The elapsed time during the braking phase is described by
\begin{multline}
\label{eq:braking_duration_state_feedback_closed_form}
\Delta t_{q_3} = - \frac{1}{b_{q_3,1}} \left( \ln \left( \frac{2 \cair v_f + u_m -  b_{q_3,1}}{2 \cair v_f + u_m + b_{q_3,1}} \right) \right. \\
+ \left. \ln \left( \frac{2 \cair v_{q_2}(t_{s_2}) + u_m + b_{q_3,1}}{2 \cair v_{q_2}(t_{s_2}) + u_m -  b_{q_3,1}} \right) \right).
\end{multline}

\bibliography{IEEEabrv,references}

\end{document}